\documentstyle[12pt]{article}
\input{epsf}
\makeatletter
\def\titlepage{\@restonecolfalse\if@twocolumn\@restonecoltrue\onecolumn
     \else \newpage \fi \thispagestyle{empty}\c@page\z@ 
    \def\thefootnote{\fnsymbol{footnote}} }
\def\endtitlepage{\if@restonecol\twocolumn \else \newpage \fi
    \def\thefootnote{\arabic{footnote}} 
    \setcounter{footnote}{0}}  
\makeatother
\newcommand{\oaftm}{T^{\sharp}M}
\newcommand{\oaftmt}{T^{\sharp}\widetilde{M}}
\newcommand{\oaxt}{\tilde{x}}
\newcommand{\oapitil}{\tilde{\pi}}
\newcommand{\oaptil}{\tilde{p}}
\newcommand{\oadgam}{\delta_{\Gamma}}
\newcommand{\oaMt}{\widetilde{M}}
\newcommand{\oaBt}{\widetilde{B}}
\newcommand{\oaHt}{\widetilde{\cal H}}
\newcommand{\oaPB}[2]{\left\{#1,#2\right\}_{\rm PB}}
\newcommand{\oamett}{\widetilde{\mbox{$ds$}}\vphantom{\mbox{$ds$}}^2}
\newcommand{\oabT}{{\bf T}}
\newcommand{\oabR}{{\bf R}}
\newcommand{\oabZ}{{\bf Z}}
\newcommand{\oahalf}{\frac{1}{2}}
\newcommand{\oachris}[3]{\Gamma_{#1}{}^{#2}{}_{#3}}
\newtheorem{oadef}[equation]{Definition}
\newtheorem{oathm}[equation]{Theorem}

\begin{document}
\begin{titlepage}
    \begin{center}
        November~1995 \hfill UMTG--188\\
        \strut\hfill {\tt hep-th/9511024}\\[1in]
        {\Large\bf Classical Geometry and Target 
            Space Duality}\footnote{Research supported in part by
            The National Science Foundation, Grant PHY--9507829.}\\[1in]
        {\bf Orlando Alvarez}\footnote{e-mail: \it 
            alvarez@phyvax.ir.miami.edu}\\[.1in]
        {\it Department of Physics\/}\\
        {\it University of Miami\/}\\
        {\it P.O. Box 248046\/}\\
        {\it Coral Gables, FL 33124 USA}\\[.5in]
\begin{abstract}
    A new formulation for a ``restricted'' type of target space 
    duality in classical two dimensional nonlinear sigma models is 
    presented.  The main idea is summarized by the analogy: euclidean 
    geometry is to riemannian geometry as toroidal target space 
    duality is to ``restricted'' target space duality.  The target 
    space is not required to possess symmetry.  These lectures only 
    discuss the local theory.  The restricted target space duality 
    problem is identified with an interesting problem in classical 
    differential geometry.
    
    These lectures were presented at the Institut d'Etudes Scientifiques 
    de Carg\`ese, 11--29~July,~1995 on Low Dimensional Applications of 
    Quantum Field Theory.
    \end{abstract}
    \end{center}
\end{titlepage}

\section{Introduction}

Target space duality is a remarkable phenomenon where different 
2-dimensional nonlinear sigma models are physically equivalent.  A key 
reason for interest in this subject is that duality often turns a 
strong coupling problem into an equivalent weak coupling one thus 
transforming an intractable problem into a manageable one.  There are 
two questions which immediately come to mind.  The difficult one is: 
given a sigma model does there exist a dual model?  A more accessible 
one is: when are two sigma models dual to each other?  In these 
lectures we will attempt to address the latter question within the 
framework of classical hamiltonian mechanics.

Basically, two sigma models are target space duals of each other if 
there exists a canonical transformation between the phase spaces which 
preserves the respective 
hamiltonians~\cite{oa-G-R-V,oa-C-Z,oa-A-AG-L2,oa-Lo}.  The existence 
of such a map is a difficult question to determine because the phase 
spaces are infinite dimensional.  To make progress, we look for 
guidance in explicit examples.  The case of toroidal target spaces 
suggests a promising approach.  The duality transformation between the 
infinite dimensional phase spaces, in the case of toroidal target 
spaces, may be viewed as being induced by a special map between some 
finite dimensional bundles over the target spaces.  In these lectures 
we address whether a similar phenomenon can arise between more general 
targets.  We will see that it is possible to look for a special type 
of ``restricted'' target space duality which leads to an intriguing 
problem in classical differential geometry.  There is a known duality 
transformation which arises when one of the target spaces is a simple 
Lie group.  The explicit form of the generating function suggests a 
generalization to more general manifolds.  This plays a pivotal role 
in our formulation.

There is a lot known about target space duality when the target is a 
torus.  The excellent review article of Giveon, Porrati and 
Rabinovici~\cite{oa-G-P-R} discusses the physics arising from toroidal 
targets in great detail.  Additionally, there are roughly 300 
references to the literature in this review which allow the reader to 
explore the historical development of the subject.  We will use 
\cite{oa-G-P-R} as our unique reference on toroidal target spaces.  The 
more recent phenomenon of ``non-abelian'' duality goes back the work 
presented in~\cite{oa-dlO-Q} but can actually be traced back to a much 
older paper~\cite{oa-F-J}.  The notion that duality can be formulated as 
canonical transformation goes back to~\cite{oa-G-R-V} even though in the 
1970's much work was done in statistical mechanics on the study of 
abelian duality in lattice systems from the partition function 
viewpoint.  The explicit construction of a canonical transformation 
including the generating function for the target space $SU(2)$ is due 
to~\cite{oa-C-Z}.

The approach taken in these lectures is different from the traditional 
approaches to duality presented in the literature (see for example the 
discussions in 
\cite{oa-dlO-Q,oa-A-AG-L1,oa-C-Z,oa-G-R2,oa-G-R3,oa-R-V}).  In all 
these approaches one has explicitly symmetries in the target space 
which play a central role in the discussion of duality.  The duality 
transformation in these theories with symmetry is some type of 
generalized Fourier transform.  I was looking for a formulation which 
did not depend on the existence of symmetries.  I wanted something 
which might be applicable to mirror symmetry~\cite{oa-yau}.  The key 
analogy to keep in mind while reading these lectures is presented in 
Table~\ref{oa-analogy}.
\begin{table}[tbp]
 \begin{center}
 \small
    \newlength{\oaoawidth} 
    \settowidth{\oaoawidth}{``restricted'' target space duality}
    \begin{tabular}{||lcl||}
        \hline\hline \parbox[t]{\oaoawidth}{\centering euclidean 
        geometry} & {\em is to} &
        \parbox[t]{\oaoawidth}{\centering riemannian geometry} 
        \\
        \multicolumn{3}{||c||}{{\sf as}}\\
        \parbox[t]{\oaoawidth}{\centering toroidal target space 
        duality} & {\em is to} & 
        \parbox[t]{\oaoawidth}{\centering ``restricted'' target 
        space duality} \\
        \hline\hline
    \end{tabular}
 \normalsize
 \end{center}
    \caption{The key analogy.}
    \protect\label{oa-analogy}
\end{table}
Mostly I will present ideas and concepts rather than detailed 
mathematical formulas.  The derivation of explicit formulas requires a 
discussion of the theory of $G$-structures, a discussion of the 
differential forms version of the Frobenius theorem, and a 
presentation of Cartan's equivalence \index{equivalence problem} 
method~\cite{oa-gardner}.  These topics are outside the scope of these 
lectures and will be presented in Part~I of \cite{oa-inprep}.  The 
ideas presented here can be generalized by weakening the requirement 
that the canonical transformation be induced by a finite dimensional 
map.  This requires the full machinery of Cartan-Kahler 
theory~\cite{oa-B-C-G3} and will be presented in Part~II of 
\cite{oa-inprep}.  These ideas can be extended to complex manifolds as 
discussed in \cite{oa-inprep}.

\section{Preliminaries}

The classical nonlinear sigma model is defined by a map $x$ from a 
lorentzian world sheet $\Sigma$ to a target manifold $M$ and some 
additional geometric data which specifies the lagrangian.  In this 
article we will take the world sheet $\Sigma$ to be either $\oabR 
\times\oabR$ or $\oabR \times S^1$.  The first factor is time and the 
second factor is space.  Local coordinates on $\Sigma$ will be denoted 
by $(\tau, \sigma)$.  The target space $M$ is endowed with a metric 
tensor $ds^2 = g_{\mu\nu} dx^\mu \otimes dx^\nu$ and a 2-form $B =
\oahalf B_{\mu\nu} dx^\mu \wedge dx^\nu$.  The 
lagrangian \index{sigma model!lagrangian} density for 
this model is
\begin{equation}
        {\cal L} = \oahalf \eta^{ab} g_{\mu\nu}(x) \partial_a x^\mu 
        \partial_b x^\nu -\oahalf \epsilon^{ab} B_{\mu\nu}(x) 
        \partial_a x^\mu \partial_b x^\nu \;, 
\end{equation}
where $\eta$ is the two dimensional lorentzian metric on the world 
sheet.  A good way of denoting the sigma model is to use the notation 
$(M,ds^2,B)$ which incorporates all the relevant geometrical data.  
The canonically conjugate momenta are given by
\begin{equation}
    \pi_\mu = {\partial {\cal L} \over \partial \dot{x}^\mu} = 
    g_{\mu\nu}(x)\dot{x}^\nu + B_{\mu\nu}(x)\frac{\partial 
    x^\nu}{\partial\sigma}\;.  
\end{equation}
Local coordinates in phase space may be taken to be 
$(x(\sigma),\pi(\sigma))$.  If the spatial part of $\Sigma$ is a 
circle then $(x(\sigma),\pi(\sigma))$ is a loop in $T^*M$, 
\index{cotangent bundle!ordinary@$T^*M$} the cotangent bundle of $M$.  
If the spatial part of $\Sigma$ is $\oabR$ then 
$(x(\sigma),\pi(\sigma))$ is a path in $T^*M$.  The symplectic form on 
the phase space is given by
\begin{equation}
    \int \delta\pi_\mu(\sigma)\wedge\delta x^\mu(\sigma)\;d\sigma\;,
    \label{oa-symplectic}
\end{equation}
where $\delta$ is the differential on the phase space.  With this 
symplectic structure we see that the basic Poisson bracket is given by
\begin{equation}
    \oaPB{x^\mu(\sigma)}{\pi_\nu(\sigma')} = \delta^\mu{}_\nu\,
        \delta(\sigma-\sigma')\;.
    \label{oa-poisson}
\end{equation}
The hamiltonian density and the worldsheet momentum density are 
respectively given by
\begin{eqnarray}
    {\cal H} & = & \oahalf g^{\mu\nu}(x)
    \left(\pi_\mu - B_{\mu\kappa}\frac{d x^\kappa}{d\sigma}\right)
    \left(\pi_\nu - B_{\nu\lambda}\frac{d x^\lambda}{d\sigma}\right)
    +\oahalf g_{\mu\nu}(x) \frac{dx^\mu}{d\sigma}\frac{dx^\nu}{d\sigma}\;,
    \label{oa-hamiltonian}\index{sigma model!hamiltonian}  \\
    {\cal P} & = & \pi_\mu \frac{d x^\mu}{d\sigma}\;.
    \index{sigma model!momentum}
\end{eqnarray}

Target space duality is the phenomenon that a nonlinear sigma model 
$(M,ds^2,B)$ is {\em equivalent\/} to a different nonlinear sigma 
model $(\oaMt,\oamett,\oaBt)$.  We need to define {\em equivalent\/}.  
For the moment we will make a preliminary definition which will be 
modified later.
\begin{oadef}[Preliminary] \label{oa-duality-prelim}
    Two sigma models $(M,ds^2,B)$ and $(\oaMt,\oamett,\oaBt)$ are said 
    to be {\em target space dual\/} to each other if there exists a 
    canonical transformation from the phase space of $(M,ds^2,B)$ to 
    the phase space of $(\oaMt,\oamett,\oaBt)$ which maps the 
    hamiltonian $\cal H$ of the first model to the hamiltonian $\oaHt$ 
    of the second model.  \index{target space 
    duality!definition!preliminary}
\end{oadef}
Later we will see that there are some important domain and range 
issues which must be addressed to have a good definition of target 
space duality.

\section{Examples}
\subsection{Circular target space}
    \index{target space!circle}
    
As an example of the above we consider the case where the target space 
is a circle of radius $R$ (for a more detailed discussion look in 
\cite{oa-G-P-R}).  The coordinate $x$ on the circle has period $2\pi R$.  
The hamiltonian density is given by
\begin{equation}
    {\cal H} = \oahalf \pi^2 + \oahalf\left(\frac{dx}{d\sigma}\right)^2\;.
\end{equation}
We can define a {\em formal\/} canonical transformation by the 
ordinary differential equations
\begin{eqnarray}
    \frac{d\oaxt}{d\sigma} & = & \pi(\sigma)\;,  
    \label{oa-can-1}\\
    \oapitil & = & \frac{dx}{d\sigma}\;. 
    \label{oa-can-2}
\end{eqnarray}
It is clear that the hamiltonian is preserved by this map.  The new 
hamiltonian tells us that the new target space is either $\oabR$ or $S^1$.

The transformation above is formal because of certain domain and range 
issues.  First we note that there is an $S^1$ action on our circle 
given by translating $x$.  Noether's theorem leads to the conserved 
target space momentum $P=\int \pi(\sigma) d\sigma$ associated with 
translations in the target space.  For the moment let us assume the 
world sheet is $\Sigma = \oabR \times S^1$.  In this case, the 
difference $x(2\pi)-x(0)$ is quantized in units of $2\pi R w$ where 
the winding number $w$ is an integer.  Consequently, phase space 
divides into sectors labeled by $[w,p_T]$ where $w$ is the winding 
number and $p_T$ is the total target space momentum.  Since both these 
quantities are conserved, the subspace labeled by $[w,p_T]$ will be 
invariant under the hamiltonian flow.  This observation has important 
consequences when we examine the duality transformation in more 
detail.  Integrating equations~(\ref{oa-can-1}) and (\ref{oa-can-2}) 
we see that:
\begin{eqnarray}
    \oaxt(2\pi)-\oaxt(0) & = & \int_{S^1} \pi(\sigma)d\sigma\;,
      \\
    \int_{S^1}\oapitil(\sigma)d\sigma & = & x(2\pi)-x(0)\;.
\end{eqnarray}
The above indicates that the dual target manifold should be a circle 
of radius $\widetilde{R}$ and we should have relations 
$2\pi\widetilde{R}\tilde{w} = p_T$ and $2\pi R w = \tilde{p_T}$.  
Since the winding number $\tilde{w}$ must be an integer we see that 
$p_T$ must be ``classically quantized'' and likewise $\tilde{p}_T$.  
At the classical level we have the following: given two radii $R$ and 
$\widetilde{R}$, the sigma model with radius $R$ on the reduced phase 
space characterized by $[w,2\pi\widetilde{R}\tilde{w}]$ is dual to the 
sigma model on a circle of radius $\widetilde{R}$ on the reduced phase 
space characterized by $[\tilde{w},2\pi R w]$.  Note that some type of 
``pre-quantization'' has taken place in trying to define duality.  
There is no ``good'' map from the full phase space of the nonlinear 
sigma model on a circle of radius $R$ to the nonlinear sigma model on 
a circle of radius $\widetilde{R}$.  Under quantization we observe 
that the momentum $P$ must be quantized in units of $1/R$, likewise, 
the momentum $\tilde{P}$ must be quantized in units of 
$1/\widetilde{R}$.  Incorporating this we see that there is now a 
relation $2\pi R\widetilde{R}=1$ between $R$ and $\widetilde{R}$.  
Note that if we take $\Sigma=\oabR\times\oabR$ then the winding number 
is not defined and the domain and range issues do not appear.  The 
lesson learned in this example is that there are delicate domain and 
range issues which must be understood if one wants to be 
mathematically precise.

Preliminary definition~(\ref{oa-duality-prelim}) 
must be expanded to include domain and range information. A better 
definition of classical duality would be
\begin{oadef} 
    Two sigma models $(M,ds^2,B)$ and $(\oaMt,\oamett,\oaBt)$ are said 
    to be {\em target space dual\/} to each other if there exists a 
    canonical transformation from a reduced phase space of 
    $(M,ds^2,B)$ to a reduced phase space of $(\oaMt,\oamett,\oaBt)$ 
    which maps the hamiltonian $\cal H$ of the first model to the 
    hamiltonian $\oaHt$ of the second model.  The reduced phase spaces 
    must be invariant under the respective hamiltonian flow.  
    \index{target space duality!definition}
\end{oadef}

\subsection{Toroidal target spaces}
    \index{target space!torus} 

In this example we choose the target space to be an $n$-dimensional 
torus $\oabT^n$ (for a more detailed discussion look in 
\cite{oa-G-P-R}).  The metric $ds^2$ and the 2-form $B$ are chosen to 
be constant.  The basic Poisson brackets for this model are given by 
equation~(\ref{oa-poisson}).  By differentiating they may be written 
as
\begin{equation}
    \oaPB{\frac{dx^\mu}{d\sigma}(\sigma)}{\pi_\nu(\sigma')} =
        \delta^\mu{}_\nu \delta'(\sigma-\sigma')\;.
\end{equation}
If we now put $dx/d\sigma$ and $\pi$ into a $2n$-vector
\begin{equation}
    z(\sigma) = 
    \left(
    \begin{array}{c}
        dx/d\sigma  \\
        \pi
    \end{array}
    \right)
\end{equation}
then the Poisson brackets may be written as
\begin{equation}
    \oaPB{z^A(\sigma)}{z^B(\sigma')} = Q^{AB} \delta'(\sigma-\sigma')\;,
\end{equation}
where
\begin{equation}
    Q = \left(
    \begin{array}{cc}
        0 & I_n  \\
        I_n & 0
    \end{array} \right)\;.
\end{equation}
The indices $A$ and $B$ range over $\{1,\ldots,2n\}$.  We see that a 
constant linear transformation $T: z \mapsto T z$ will preserve the 
symplectic structure if $T$ is in the pseudo-orthogonal group 
$O_Q(2n)$ \index{porth@$O_Q(2n)$} consisting of linear transformations 
which preserve the quadratic form $Q$.  This group is isomorphic to 
$O(n,n)$.  Since the hamiltonian is a quadratic form in the $z$'s with 
constant coefficients, we see that a $T\in O_Q(2n)$ leads to a new 
sigma model hamiltonian with constant coefficients.  A similar 
phenomenon will be studied in detail in a more general setting later.  
For the moment we make a few observations.  Since the hamiltonian a 
positive definite quadratic form in $\{dx/d\sigma,\pi\}$, the linear 
transformations in $O_Q(2n)$ that preserve the quadratic form belong 
to a certain maximal compact subgroup $K$ which is isomorphic to 
$O(n)\times O(n)$.  The dimension of the coset space $O_Q(2n)/K$ is 
$n^2$ which is precisely the total number of independent components in 
the metric $ds^2$ and the 2-form $B$.  In fact the coset space 
$O_Q(2n)/K$ parameterizes the space of nonlinear sigma model 
hamiltonians with constant coefficients.  Actually one has to be 
careful in the quantum theory.  One can show that transformations in 
the subgroup $O(n,n;\oabZ)$ lead to equivalent hamiltonians.

In this report we are interested in local conditions which necessarily 
guarantee the existence of a restricted type of target space duality.  
Because we are only interested in local issues we will generally take 
$\Sigma=\oabR\times\oabR$.

\subsection{Toroidal target spaces revisited}

We revisit toroidal target spaces and adopt a different viewpoint 
which will generalize to generic targets.  There is a very interesting 
structure which arises in the models we have been studying.  Introduce 
a new space isomorphic to $\oabR^n$ with coordinates $p$.  Define 
$p(\sigma)$ by
\begin{equation}
    \frac{dp}{d\sigma}= \pi(\sigma)\;.
    \label{oa-p-def-prelim}
\end{equation}
Note that there is an ambiguity in the definition of $p$ due to the 
constant of integration.  Instead of working in phase space 
$(x(\sigma),\pi(\sigma))$ we can work in a space with coordinates 
$(x(\sigma),p(\sigma))$. In terms of these new variables the 
symplectic form may be written as
\begin{equation}
    \label{oa-symplectic-2}
    \int \delta x^\mu(\sigma) \wedge \frac{d}{d\sigma}\,
    \delta p_\mu(\sigma) \; d\sigma
    = \oahalf \int d\sigma\;
    \left(
    \begin{array}{cc}
        \delta x & \delta p
    \end{array}
    \right)
    \wedge \frac{d}{d\sigma}
    \left(
    \begin{array}{cc}
        0 & I_n  \\
        I_n & 0
    \end{array}
    \right)
    \left(
    \begin{array}{c}
        \delta x  \\
        \delta p
    \end{array}
    \right) \;.
\end{equation}
This symplectic form is degenerate.  We have also discarded a surface 
term.  We will ignore these technical issues.  Note that the matrix 
$Q$ enters into this formulation of the symplectic form.  Also, a 
constant coefficient linear transformation $T\in O_Q(2n)$ acting on 
$(x,p)$ space will preserve this symplectic form.  Using the variables 
$(x,p)$ we can integrate equations~(\ref{oa-can-1}), (\ref{oa-can-2}), and 
obtain
\begin{eqnarray}
    \tilde{x}(\sigma) & = & p(\sigma) + a \;,
    \label{oa-int-1}  \\
    \tilde{p}(\sigma) & = & x(\sigma) + b\;,
    \label{oa-int-2}
\end{eqnarray}
where $a$ and $b$ are constants.  In passing we mention that if 
$\Sigma= \oabR \times S^1$ and if we naively quantize then there will be 
appropriate periodicity requirements on $(x,p)$ and 
$(\tilde{x},\tilde{p})$.  The map $(x,p) \mapsto (\oaxt=p+a,\oaptil=x+b)$ 
between the toroidal target spaces $\oabT^{2n}$ induces transformations 
(\ref{oa-int-1}) and (\ref{oa-int-2}) on the space of paths.

The question which will occupy our attention throughout this report 
is whether it is possible to formulate target space duality as an 
induced map between the infinite dimensional phase spaces arising from 
an ordinary map from a $2n$-manifold to another $2n$-manifold.

\begin{figure}[tbp]
    \centerline{\epsfxsize 0.6\textwidth \epsfbox{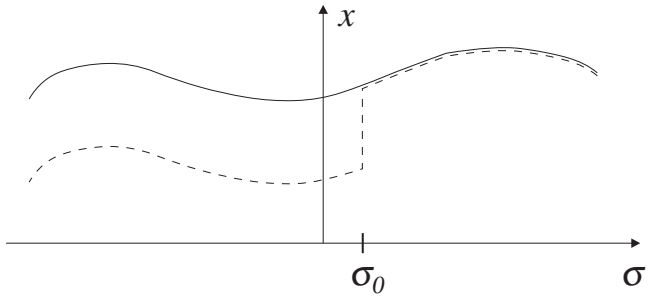}} 
    \caption[kink]{The action of $\exp[2\pi iR p(\sigma_0)]$ takes a 
    state with expectation value $x_{\rm cl}(\sigma)$ (the solid 
    curve) to a state with expectation value $x_{\rm 
    cl,\,new}(\sigma)$ (the dashed curve).}
    \protect\label{oa-fig-kink}
\end{figure}

The variables $p$ have an important physical significance.  They are 
the variables which describe the ``solitons'' in toroidal models.  It 
is best to take a semi-classical viewpoint.  By the canonical 
commutation relations we have that
\begin{equation}
    \pi_\mu(\sigma) = -i \frac{\delta}{\delta x^\mu(\sigma)}\;.
\end{equation}
Therefore we have
\begin{equation}
    \exp[i\alpha^\mu p_\mu(\sigma)] = \exp\left( \alpha^\mu 
    \int_{-\infty}^\sigma d\sigma'\; \frac{\delta}{\delta 
    x^\mu(\sigma')} \right)\;.
\end{equation}
This equation tells us that $\exp[i\alpha\cdot p(\sigma)]$ is the operator 
which creates a ``kink'' with jump of size $\alpha$.  To see this let 
us assume the target space is a circle of radius $R$ and consider a 
state with the property that $\langle x(\sigma)\rangle = x_{\rm 
cl}(\sigma)$.  The action of the operator $\exp[2\pi i R p(\sigma_0)]$ 
on the state leads to a new state with expectation value $x_{\rm 
cl,\,new}(\sigma)$ which contains a $2\pi R$ kink at $\sigma_0$ as in 
Figure~\ref{oa-fig-kink}.

In the toroidal models we see that the space with coordinates $(x,p)$ 
is a space which describes both the ``particles'', {\em i.e.}, the 
$x$'s, and the ``solitons'', {\em i.e.}, the $p$'s.  The torus 
$\oabT^{2n}$ is determined by a lattice in $\oabR^{2n}$.  The group 
$O_Q(2n)$ naturally acts on this $\oabR^{2n}$.  The lattice will be 
invariant under $O(n,n;\oabZ) \subset O_Q(2n)$.  This action preserves 
the symplectic form (\ref{oa-symplectic-2}).

\section{Generating functions}
    \index{generating function}

The term {\em generating function} is used in a variety of different 
contexts in mechanics.  Assume we have a hamiltonian system.  The 
equations of motion may be derived as the extremals of the variational 
principle defined by the action $I=\int (pdq - H(q,p) dt)$.  Assume we 
have a second system with canonical coordinates 
$(\tilde{q},\tilde{p})$, time independent hamiltonian 
$\widetilde{H}(\tilde{q},\tilde{p})$ and action $\tilde{I}$.  The 
variational principle for $I$ is equivalent to the variational 
principle for $\tilde{I}$ if $pdq - H dt$ differs from 
$\tilde{p}d\tilde{q}-\widetilde{H} dt$ by a total differential:
\begin{equation}
    \tilde{p}d\tilde{q}-\widetilde{H} dt =
    pdq - H dt + dG\;.
\end{equation}
If the function $G$ is time independent then
\begin{eqnarray}
    \tilde{p} & = & \frac{\partial G}{\partial \tilde{q}}\;,
     \\
    p & = & - \frac{\partial G}{\partial q}\;,
\end{eqnarray}
and $H = \widetilde{H}$.  The function $G(q,\tilde{q})$ is called the 
{\em generating function\/}.  A geometrical discussion from the 
viewpoint of symplectic geometry may be found in \cite{oa-A-M,oa-Ar}.

For example, in the simple harmonic oscillator with $H 
=\oahalf(p^2+q^2)$, the generating function $G(q,\tilde{q})=q\tilde{q}$ 
leads to the ``duality transformation'' $\tilde{p} = q$ and $p = 
-\tilde{q}$.

In a field theoretic context we have that for a circular target space 
the duality transformation is generated by the functional
\begin{equation}
    G[x,\tilde{x}] = \int \tilde{x}(\sigma)\,
    \frac{dx}{d\sigma}\,d\sigma \;.
\end{equation}
A simple computation shows that 
\begin{eqnarray*}
    \oapitil & = & \frac{\delta G}{\delta \tilde{x}}=\frac{dx}{d\sigma}\;,\\
    \pi & = & -\frac{\delta G}{\delta x} = \frac{d\tilde{x}}{d\sigma}\;.
\end{eqnarray*}
Note that the integrand is the pullback of a 1-form on the product 
space $S^1 \times S^1$ of the two target circles, and that $G$ is 
reparametrization invariant.

\section{Generic target space}

\subsection{Background}

We now address whether it is possible to construct some type of theory 
which addresses duality in a generic sigma model $(M,ds^2,B)$.  In our 
construction, the group $O_Q(2n)$ will appear but in a different 
manner.  The reader is reminded of the analogy presented in 
Table~\ref{oa-analogy}.  In these notes we will address a certain 
``restricted'' type of duality which leads to a well-posed 
mathematical problem.  The more general discussion requires the use of 
the full machinery of exterior differential systems and is beyond the 
scope of these lectures~\cite{oa-inprep}.  Remember that the circular 
duality transformation may viewed as an induced map on paths arising 
from an affine map from $(x,p)$-space to $(\oaxt,\oaptil)$-space.  In 
this section we address whether it is possible to have a similar 
phenomenon arise when we have a generic target space.  We will see 
that in certain situations duality can arise as an ordinary map from a 
certain $2n$-manifold to another $2n$-manifold.

The starting point of our discussion will be generating functions.  
Assume we have two sigma models $(M,ds^2,B)$ and 
$(\oaMt,\oamett,\oaBt)$.  We would like to know if there is duality 
transformation between these models.  Let us postulate that the 
generating function has the form
\begin{equation}
    G[x,\tilde{x}] = \int \left( 
    u_\mu(x(\sigma),\tilde{x}(\sigma)) \frac{dx^\mu}{d\sigma} +
    v_\mu(x(\sigma),\tilde{x}(\sigma))\frac{d\tilde{x}^\mu}{d\sigma}
    \right) d\sigma \;.
    \label{oa-gen-fcn}
\end{equation}
This generating function is a generalization to arbitrary manifolds of 
the one postulated for $SU(2)$ in~\cite{oa-C-Z}.  Such a generating 
function has many desirable properties.  It is reparametrization 
invariant.  The integrand is the pullback of the 1-form 
$u_\mu(x,\tilde{x}) dx^\mu + v_\mu(x,\tilde{x}) d\tilde{x}^\mu$ on 
$M\times\oaMt$.  Since we are interested in theories where the dynamical 
variables are paths on a manifold we see that (\ref{oa-gen-fcn}) is very 
natural.  It is simply the integral of a 1-form along a curve in the 
product manifold $M\times\oaMt$.  A brief computation shows that
\begin{eqnarray*}
    \pi_\mu(\sigma) & = & 
    a_{\mu\nu}(x(\sigma),\tilde{x}(\sigma)) \frac{dx^\nu}{d\sigma} 
    + b_{\mu\nu}(x(\sigma),\tilde{x}(\sigma)) 
    \frac{d\tilde{x}^\nu}{d\sigma}\;,
    \\
    \tilde{\pi}_\mu(\sigma) & = & 
    c_{\mu\nu}(x(\sigma),\tilde{x}(\sigma)) \frac{dx^\nu}{d\sigma} 
    + d_{\mu\nu}(x(\sigma),\tilde{x}(\sigma)) 
    \frac{d\tilde{x}^\nu}{d\sigma}\;. 
\end{eqnarray*}
Solving for the domain and range variables we get
\begin{eqnarray}
    \frac{d\tilde{x}^\mu}{d\sigma} & = & 
    A^\mu{}_\nu(x(\sigma),\tilde{x}(\sigma)) \frac{dx^\nu}{d\sigma}
    +
    B^{\mu\nu}(x(\sigma),\tilde{x}(\sigma)) \pi_\nu(\sigma)\;,
    \label{oa-c-1}  \\
    \tilde{\pi}_\mu(\sigma) & = & 
    C_{\mu\nu}(x(\sigma),\tilde{x}(\sigma)) \frac{dx^\nu}{d\sigma}
    +
    D_\mu{}^\nu(x(\sigma),\tilde{x}(\sigma)) \pi_\nu(\sigma)\;.
    \label{oa-c-2}
\end{eqnarray}
This canonical transformation linear in $dx/d\sigma$ and $\pi$ is very 
suggestive of target space duality.  The reason is that the 
hamiltonian is quadratic in $dx/d\sigma$ and $\pi$ and thus gets 
mapped into something which is quadratic in $d\tilde{x}/d\sigma$ and 
$\tilde{\pi}$.  The difficulty is that the new metric and 
anti-symmetric tensor might not be functions of only $\tilde{x}$.  
{\em This is the obstruction to duality under 
ansatz}~(\ref{oa-gen-fcn}).  Part of our discussion will be to try to 
understand whether there are local obstructions to the existence of 
duality transformations.  It is worthwhile stating this again 
explicitly.
\begin{quote}
    Do canonical transformations (\ref{oa-c-1}) and (\ref{oa-c-2}) 
    lead to a new $\widetilde{\cal H}$ such that $\oamett$ and $\oaBt$ 
    are only functions of $\tilde{x}$?  In general $\widetilde{\cal 
    H}$ will be a non-local function of the variables.
\end{quote}
Finally, a more complicated ansatz for (\ref{oa-gen-fcn}) which might 
include terms such as $d^2x/d\sigma^2$ or $\sqrt{(dx/d\sigma)^2}$ will 
generally not lead to transformation equations which are linear in 
$dx/d\sigma$ and $\pi$ as exemplified in (\ref{oa-c-1}) and 
(\ref{oa-c-2}).

We proceed to attack these issues.  Note that equations (\ref{oa-c-1}) 
and (\ref{oa-c-2}), being ordinary differential equations (ODE), are 
always integrable.  To explicitly see this, assume we are given 
$(x(\sigma),\pi(\sigma))$ then we insert them into 
equation~(\ref{oa-c-1}) which we integrate\footnote{Of course there 
are arbitrary constants of integration.} to obtain 
$\tilde{x}(\sigma)$.  Subsequently we insert 
$x(\sigma),\pi(\sigma),\tilde{x}(\sigma)$ into (\ref{oa-c-2}) to get 
$\oapitil(\sigma)$.  What we are going to do is to replace equations 
(\ref{oa-c-1}) and (\ref{oa-c-2}) by an equivalent set of equations 
(up to ambiguities involving constants of integration).

Introduce new variables $p$ and $\oaptil$ and define $p(\sigma)$ and 
$\oaptil(\sigma)$ by
\begin{eqnarray}
    \frac{dp}{d\sigma} & = & \pi(\sigma)\;,
     \\
    \frac{d\oaptil}{d\sigma} & = & \oapitil(\sigma)\;.
\end{eqnarray}
It can be shown~\cite{oa-inprep} that the space with coordinates 
$(x,p)$ is a manifold $\oaftm$ 
\index{cotangent bundle!fake@$T^{\sharp}M$} 
isomorphic\footnote{In the case of toroidal 
target spaces we saw that $(x,p)$ space was a $2n$-torus rather than 
$T^*(\oabT^n)$ due to some global and quantum properties.  All our 
discussions are local thus we will not see such phenomena.} to $T^*M$.  
I do not understand the relationship between the use
of the cotangent bundle here and in the interesting work  
presented in
\cite{oa-K-S1,oa-K-S2}.  Canonical transformation equations 
(\ref{oa-c-1}) and (\ref{oa-c-2}) may be rewritten as
\begin{eqnarray}
    \frac{d\tilde{x}^\mu}{d\sigma} & = & 
    A^\mu{}_\nu(x(\sigma),\tilde{x}(\sigma)) \frac{dx^\nu}{d\sigma}
    +
    B^{\mu\nu}(x(\sigma),\tilde{x}(\sigma)) \frac{dp_\nu}{d\sigma}\;,
    \label{oa-xp-1}  \\
    \frac{d\oaptil_\mu}{d\sigma} & = & 
    C_{\mu\nu}(x(\sigma),\tilde{x}(\sigma)) \frac{dx^\nu}{d\sigma}
    +
    D_\mu{}^\nu(x(\sigma),\tilde{x}(\sigma)) \frac{dp_\nu}{d\sigma}\;.
    \label{oa-xp-2}
\end{eqnarray}
These are integrable because they are ODE's. An advantage of the above 
over (\ref{oa-c-1}) and (\ref{oa-c-2}) is that the above treat $x$ and $p$ 
symmetrically.

Is is worthwhile to make a small mathematical digression.  Given a 
manifold $N$, let $PN=\{\gamma:\oabR \to N\}$ be the space of all 
paths in $N$.  Every map $f:N\to \tilde{N}$ induces a map $f_{*}:PN\to 
P\tilde{N}$.  The opposite is not true.  Not all maps from $PN$ to 
$P\tilde{N}$ arise as maps from $N$ to $\tilde{N}$.  This is 
illustrated in equation (\ref{oa-xp-1}) which is a map from 
$P(\oaftm)$ to $P(\oaftmt)$.  Here we explicitly see that the map 
depends not only on the point $(x,\oaxt)\in \oaftm\times \oaftmt$ but 
also on the ``velocities'' $dx/d\sigma$ and $dp/d\sigma$.  In general 
equations (\ref{oa-xp-1}) and (\ref{oa-xp-2}) do not define a map from 
$\oaftm$ to $\oaftmt$.  We are interested in looking for a phenomenon 
similar to that discussed after equation~(\ref{oa-symplectic-2}) where 
the map between the phase spaces is induced by a map between $\oaftm$ 
and $\oaftmt$.

In these lectures we address the question of when do equations 
(\ref{oa-xp-1}) and (\ref{oa-xp-2}) arise as maps from $\oaftm$ to 
$\oaftmt$ in such a way that we get a dual sigma model.  We proceed 
naively by performing a formal mathematical manipulation on 
(\ref{oa-xp-1}) and (\ref{oa-xp-2}): let us multiply both sides of the 
equations by $d\sigma$ and obtain the {\em exterior differential 
system} (EDS) which will schematically be written as
\begin{eqnarray}
    d\oaxt & = & A dx + B dp\;,
    \label{oa-d-1}  \\
    d\oaptil & = & C dx + D dp\;.
    \label{oa-d-2}
\end{eqnarray}
This exterior differentials system is equivalent to the partial 
differential equations (PDE) $\partial\oaxt/\partial x = A$, 
$\partial\oaxt/\partial p = B$, $\partial\oaptil/\partial x = C$ and 
$\partial\oaptil/\partial p = D$.  In general PDE's have no solutions.  
There are certain integrability conditions which must be satisfied in 
order for the system to be integrable.  Roughly, we have to be able to 
integrate simultaneously along $2n$ independent directions.  This is 
very different from our original ODE system which has no integrability 
conditions.  System (\ref{oa-xp-1}) and (\ref{oa-xp-2}) is equivalent 
to finding one dimensional integrable manifolds of (\ref{oa-d-1}) and 
(\ref{oa-d-2}).  What we are proposing is that we look for 
$2n$-dimensional integrable manifolds of (\ref{oa-d-1}) and 
(\ref{oa-d-2}).  If such an integrable manifold exists then we have a 
map from $\oaftm$ to $\oaftmt$.  The integrability conditions for EDS 
(\ref{oa-d-1}) and (\ref{oa-d-2}) will describe local geometric 
conditions which must be satisfied on $\oaftm$ and $\oaftmt$ in order 
for a solution to exist.

\subsection{Details}

We now turn to more detailed study of what was proposed in the last 
section.  Assume we are given a sigma model $(M,ds^2,B)$.  Let 
$\Gamma$ be the Levi-Civita connection associated with metric $ds^2$.  
We introduce new variables $p$ and relate them to $\pi$ via
\begin{equation}
    \pi_\mu = \frac{dp_\mu}{d\sigma} 
    - \frac{dx^\lambda}{d\sigma}\oachris{\lambda}{\nu}{\mu} p_\nu
    = \frac{\nabla}{d\sigma} p_\mu\;.
\end{equation}
This definition is different than the one previously given by 
equation~(\ref{oa-p-def-prelim}).  The definition given above is much 
better because $p$ transforms covariantly with respect to coordinate 
transformations on the base space $M$.  Just as before one can show 
that the space with coordinates $(x,p)$ is a manifold $\oaftm$ 
isomorphic to the cotangent bundle $T^*M$.  For any covariant vector 
$v_\mu$ define the covariant variation $\oadgam v$ by
\begin{equation}
    \oadgam v_\mu = \delta v_\mu - \delta x^\lambda 
    \oachris{\lambda}{\nu}{\mu} v_\nu \;.
\end{equation}
A brief computation shows that 
\begin{equation}
    \oadgam \pi_\mu = \frac{\nabla}{d\sigma} \oadgam p_\mu
    - R^\nu{}_{\mu\lambda\rho} \delta x^\lambda \frac{dx^\rho}{d\sigma} 
    p_\nu \;.
\end{equation}

The symplectic structure 
(\ref{oa-symplectic}) may be rewritten as
\begin{eqnarray}
    && \int d\sigma \; \delta x^\mu \wedge \delta \pi_\mu
     \nonumber\\
     &=&
    \oahalf \int d\sigma\;
    \left(
    \begin{array}{cc}
        \delta x & \oadgam p
    \end{array}
    \right)
    \wedge 
    \left(
    \begin{array}{cc}
        0 & I_n  \\
        I_n & 0
    \end{array}
    \right)
    \left(
    \begin{array}{cc}
        \nabla/d\sigma & 0  \\
        b & \nabla/d\sigma
    \end{array}
    \right)
    \left(
    \begin{array}{c}
        \delta x  \\
        \oadgam p
    \end{array}
    \right) \;,
    \label{oa-symplectic-f}
\end{eqnarray}
where $b_{\mu\nu} = p_\lambda (dx^\rho/d\sigma) 
R^\lambda{}_{\rho\mu\nu}$. Note that our old friend $Q$ appears in 
the above. What is surprising is that
\begin{equation}
    \left(
    \begin{array}{cc}
        \nabla/d\sigma & 0  \\
        b & \nabla/d\sigma
    \end{array}
    \right)
    \label{oa-p-conn}
\end{equation}
is the unique torsion free pseudo-riemannian connection on $\oaftm$ 
associated with the $Q$-metric \index{metric!qmetric@$ds^2_Q$}
\begin{equation}
    ds^2_Q = dx^\mu \otimes (dp_\mu - dx^\lambda 
    \oachris{\lambda}{\nu}{\mu} p_\nu) + 
    (dp_\mu - dx^\lambda \oachris{\lambda}{\nu}{\mu} p_\nu)
    \otimes dx^\mu
\end{equation}
on $\oaftm$.  A short computation shows that (\ref{oa-p-conn}) is a 
skew-adjoint operator with respect to the $Q$-metric.

We see that some natural geometric structures related to the 
pseudo-riemannian geometry of the $Q$-metric are beginning to appear.

\subsection{A digression and an analogous problem}

We expand on the analogy discussed in Table~\ref{oa-analogy}.  
Assume we are in euclidean space $\oabR^n$.  The euclidean group $E(n)$ 
is the group of isometries of euclidean space.  Since Felix Klein 
\index{Klein, Felix} we have understood that euclidean geometry is the 
study of the properties of figures which are invariant under the 
action of the euclidean group (the isometry group of euclidean space).  
In a similar fashion one can define hyperbolic and elliptic 
geometries.  Klein's ideas seem to fail when one considers riemannian 
geometry: on a generic riemannian manifold $M$ there are no 
isometries.  E.~Cartan \index{Cartan, Elie} realized that this was not 
fatal and that the orthogonal group played a very important role in 
riemannian geometry.  Cartan reformulated euclidean geometry using the 
observation that $\oabR^n = E(n)/O(n)$.  In modern language, $E(n)$ is 
a principal $O(n)$-bundle over $\oabR^n$.  Cartan studied the 
properties of $\oabR^n$ in terms of the properties of $E(n)$.  In 
doing so he realized that the Maurer-Cartan equations for the group 
$E(n)$ contain all the information necessary to extract both the 
properties of $\oabR^n$ and $O(n)$.  Cartan now attacked the problem 
of riemannian geometry by observing that on every riemannian manifold 
$(M,ds^2)$ one could construct a larger space $O(M)$ called the bundle 
of orthonormal frames.  For a euclidean space, the bundle of 
orthonormal frames may be identified with $E(n)$.  Cartan was able to 
write down a generalization of the Maurer-Cartan equations on the 
bundle $O(M)$.  These equation are known as the first and second 
structural equations of the space.  What Cartan discovered was that in 
riemannian geometry there was a group action of $O(n)$ on the bundle 
of frames $O(M)$ rather than on the base space $M$.  The base space $M 
= O(M)/O(n)$ and all properties of $M$ may be understood in terms of 
the properties of $O(M)$.  This is the starting point for modern 
riemannian geometry.

Assume we have an isometry \index{isometry!classical view} from a 
riemannian manifold $(M,ds^2)$ to a riemannian manifold $(\oaMt,\oamett)$.  
This leads to a system of non-linear PDE's given by
\begin{equation}
    \tilde{g}_{\mu\nu}(\oaxt) \frac{\partial \oaxt^\mu}{\partial x^\rho}
    \frac{\partial \oaxt^\nu}{\partial x^\sigma} = g_{\rho\sigma}(x)\;.
    \label{oa-isom}
\end{equation}
Cartan realized that there was a better and more geometric way to 
formulate these equations.  Cartan constructed local orthonormal 
coframes $\omega^m = e^m{}_\mu dx^\mu$ on $M$, and $\tilde{\omega}$ on 
$\oaMt$ where the metrics may be written as $ds^2 = \omega^m \otimes 
\omega^m$ and $\oamett = \tilde{\omega}^m \otimes \tilde{\omega}^m$.  An 
isometry requires the existence of an orthogonal matrix-valued 
function $R$ such that $\tilde{\omega}^m = R^m{}_n \omega^n$.  Cartan 
now observed that this leads us to a first order EDS 
$\tilde{e}^m{}_\mu d\oaxt^\mu = R^n{}_m e^n{}_\nu dx^\nu$.  Cartan went 
much further.  He realized that one could promote $R$ to a new 
independent variable.  This is similar to introducing a new variable 
$u=dy/dx$ in a second order ODE and writing the original equation as a 
pair of first order ODEs.  Instead of working on $M$, Cartan worked on 
a space with coordinates $(x,R)$.  In modern language, this is the 
bundle of orthonormal frames.  Also, the structural equations are 
globally defined on the bundle of orthonormal frames whereas 
(\ref{oa-isom}) are only valid locally.  This is what led to his 
invention of the theory of $G$-structures (principal sub-bundles of 
the frame bundle) and generalized geometries.

In the modern viewpoint, an isometry \index{isometry!bundle view} may 
be defined in the following way.  Assume we have a map $f:M \to 
\oaMt$; the differential of this map $df: TM \to T\oaMt$ naturally 
acts on the bundle of {\em all} frames, {\em i.e.}, 
$\frac{\partial}{\partial x^\nu} = \frac{\partial \oaxt^\mu}{\partial 
x^\nu} \frac{\partial}{\partial \oaxt^\mu}$.  If $df$ lifts to a 
bundle map from the {\em orthonormal frame bundle} $O(M)$ to the {\em 
orthonormal frame bundle} $O(\oaMt)$ then $f$ is an isometry.  We have 
to be careful with the converse.  A bundle map $O(M)\to O(\oaMt)$ is 
fiber-preserving and thus induces a map $M\to\oaMt$ on the base 
spaces.  In general, the bundle map from $O(M)$ to $O(\oaMt)$ will not 
be an isometry because it will not be the lift of a map between the 
bases.
\begin{quote}
    When is a bundle map $F: O(M)\to O(\oaMt)$ the lift of a map $f: 
    M\to\oaMt$?  Recall that frame bundles are endowed with a globally 
    defined $\oabR^n$-valued canonical 1-form.  Let $\theta$ and 
    $\tilde{\theta}$ be the respective canonical 1-forms on $O(M)$ 
    and $O(\oaMt)$.  If  $F^*\tilde{\theta}=\theta$ then $F$ is the 
    lift of a map between the bases.  This means that an isometry can 
    be defined as a bundle map between orthogonal frame bundles which 
    preserves the canonical 1-forms.
\end{quote}
The formulation just presented is global whereas the formulation in 
terms of PDE's (\ref{oa-isom}) is local.  The question of the existence 
of an isometry between riemannian manifolds is a difficult one.  There 
are both local and global issues which must be addressed.  In this 
report we only discuss local issues.  Roughly, the local existence of 
an isometry is guaranteed if there exists coordinate systems on $M$ 
and $\oaMt$ such that the curvature and its higher derivatives agree up 
to a certain order.  Global issues are much more 
difficult~\cite{oa-gromov}.

The key point is contained in the following scenario.
\begin{quote}
    Assume we have a pseudo-isometry from $(\oaftm,ds^2_Q)$ to 
    $(\oaftm,\oamett_Q)$.  It is relatively easy to show that not only 
    does the pseudo-isometry preserve the metric but also the 
    connection.  Consequently, the induced map on paths preserves the 
    symplectic form (\ref{oa-symplectic-f}).  This means that the 
    pseudo-isometry between the finite dimensional spaces $\oaftm$ and 
    $\oaftmt$ induces a canonical transformation from the infinite 
    dimensional phase space $(x(\sigma),\pi(\sigma))$ to the infinite 
    dimensional phase space $(\oaxt(\sigma),\oapitil(\sigma))$.
\end{quote}
The above requires some qualifications because we have ignored global 
issues\footnote{A discussion of global issues in duality may be found 
in~\cite{oa-A-AG-B-L}.}.  The construction just described justifies 
the analogy presented in Table~\ref{oa-analogy}.

\subsection{Back to duality}

We will now perform the mathematical construction we require.  We 
begin with a sigma model $(M,ds^2,B)$.  Using the metric we construct 
a local orthonormal coframe $\{\omega^m\}$.  We put coordinates on the 
``fake cotangent bundle'' $\oaftm$ by noting that a 1-form may be 
written as $p_m \omega^m$.  This allows us to define 1-forms $\mu_m = 
dp_m - \omega^l \oachris{l}{n}{m}p_n$.  These 1-forms define the 
horizontal distribution on $\oaftm$ associated with the Levi-Civita 
connection.  The $Q$-metric on $\oaftm$ is given by
\begin{equation}
    ds^2_Q = \omega^m \otimes \mu_m + \mu_m\otimes\omega^m\;.
\end{equation}
A pseudo-isometry is a map from $\oaftm$ to $\oaftmt$ such that 
\begin{equation}
    \left(
    \begin{array}{c}
        \tilde{\omega}  \\
        \tilde{\mu}
    \end{array}
    \right) =
    S \left(
    \begin{array}{c}
        \omega  \\
        \mu
    \end{array}
    \right) \;,
\end{equation}
where $S \in O_Q(2n)$. We now rephrase the above in the language of 
bundles.

Given the metric $ds^2_Q$ on $\oaftm$ we construct the bundle of 
$Q$-orthonormal frames, denoted by $O_Q(\oaftm)$.  This bundle has 
base space $\oaftm$ and fiber $O_Q(2n)$.  A pseudo-isometry between 
$(\oaftm,ds^2_Q)$ and $(\oaftmt,\oamett_Q)$ is given by a bundle map 
between $O_Q(\oaftm)$ and $O_Q(\oaftmt)$ which preserves the canonical 
1-forms.  Once we have a pseudo-isometry we can construct the desired 
canonical transformation.  This is the full story as far as the 
``canonical structure'' is concerned.  This is only half the problem 
because we have to worry about the hamiltonian.

\section{The hamiltonian structure}

The hamiltonian is related to a positive definite metric 
\index{metric!hmetric@$ds^2_{\cal H}$} on $\oaftm$ defined by
\begin{equation}
    ds^2_{\cal H} = (\mu_m - B_{mn}\omega^n) \otimes
    (\mu_m - B_{ml}\omega^l) + \omega^m \otimes \omega^m\;.
\end{equation}
Note that for any curve $(x(\sigma),p(\sigma))$ on $\oaftm$, the 
evaluation of the above on the tangent vector to the curve yields 
$2{\cal H}$ where $\cal H$ is defined by (\ref{oa-hamiltonian}).

We are now in a position to see what is required to have duality 
within our scenario.  Assume we have a $Q$-pseudo-isometry 
$f:\oaftm\to\oaftmt$.  If in addition, $f$ preserves the metric 
$ds^2_{\cal H}$ then the sigma model $(M,ds^2,B)$ will be dual to the 
sigma model $(\oaMt,\oamett,\oaBt)$.  We are interested in maps from 
$\oaftm$ to $\oaftmt$ which preserve two different symmetric 
2-tensors, $ds^2_Q$ and $ds^2_{\cal H}$, of different type.  We will 
call such maps $K$-isometries.  \index{isometry!$K$-isometry} To 
formulate this problem in terms of bundles we have to use the bundle 
consisting of frames which are simultaneously orthonormal with respect 
to $ds^2_Q$ and $ds^2_{\cal H}$.  The fiber of this bundle is 
isomorphic to $K = O_Q(2n) \cap O(2n) \approx O(n)\times O(n)$, the 
maximal compact subgroup of $O_Q(2n)$.  We can now state the main 
result of these lectures.  Let $K(\oaftm,ds^2_Q,ds^2_{\cal H})$ be the 
bundle consisting of frames which are simultaneously orthonormal with 
respect to $ds^2_Q$ and $ds^2_{\cal H}$.
\begin{oathm}
    Assume there exists a bundle map between the frame bundles 
    $K(\oaftm,ds^2_Q,ds^2_{\cal H})$ and 
    $K(\oaftmt,\oamett_Q,\oamett_{\widetilde{\cal H}})$ which 
    preserves the canonical 1-forms; then the sigma model $(M,ds^2,B)$ 
    is dual to the sigma model $(\oaMt,\oamett,\oaBt)$.  Said 
    differently, the sigma model $(M,ds^2,B)$ is dual to the sigma 
    model $(\oaMt,\oamett,\oaBt)$ if there exists a $K$-isometry 
    between $\oaftm$ and $\oaftmt$.
\end{oathm}

To relate these ideas to concepts familiar from toroidal target space 
duality we discuss in more detail the bundle 
$K(\oaftm,ds^2_Q,ds^2_{\cal H})$.  A general $Q$-orthonormal frame 
{\em at a point} in $\oaftm$ may be written as
\begin{equation}
    S \left(
    \begin{array}{c}
        \omega  \\
        \mu
    \end{array}
    \right)
    \label{oa-gen-frame}
\end{equation}
where $S\in O_Q(2n)$.  First we decompose $S$ in a way that 
explicitly exhibits the hamiltonian.  The Lie algebra of $O_Q(2n)$ may 
be written in a block decomposition as
\begin{equation}
    \left(
    \begin{array}{cc}
        \delta & \beta  \\
        \gamma & -\delta^t
    \end{array}
    \right)\;,
\end{equation}
each entry is an $n\times n$ matrix and $\beta,\gamma$ are skew. A 
matrix in the Lie algebra of $K$ is of the form
\begin{equation}
    \left(
    \begin{array}{cc}
        \alpha & \beta  \\
        \beta & \alpha
    \end{array}
    \right)\;,
\end{equation}
where $\alpha,\beta$ are skew. This suggests a decomposition
\begin{equation}
    \left(
    \begin{array}{cc}
        \delta & \beta  \\
        \gamma & -\delta^t
    \end{array}
    \right)
    =
    \left(
    \begin{array}{cc}
        \alpha & \beta  \\
        \beta & \alpha
    \end{array}
    \right)
    +
    \left(
    \begin{array}{cc}
        \sigma & 0  \\
        0 & -\sigma
    \end{array}
    \right)
    +
    \left(
    \begin{array}{cc}
        0 & 0  \\
        \epsilon & 0
    \end{array}
    \right)\;,
\end{equation}
where $\sigma$ is symmetric and $\epsilon$ is skew.  Near the identity 
one can rewrite~(\ref{oa-gen-frame}) in the form
\begin{equation}
    T
    \left(
    \begin{array}{cc}
        e^\sigma & 0  \\
        0 & e^{-\sigma}
    \end{array}
    \right)
    \left(
    \begin{array}{cc}
        I & 0  \\
        -B & I
    \end{array}
    \right)
    \left(
    \begin{array}{c}
        \omega  \\
        \mu
    \end{array}
    \right) 
    =
    T
    \left(
    \begin{array}{c}
         e^{+\sigma} \omega \\
        e^{-\sigma}(\mu - B\omega)
    \end{array}
    \right) \;,
    \label{oa-decomp-c}
\end{equation}
where $T\in K$ and $B$ is skew.  Note that $e^{\pm\sigma}$ are 
non-singular symmetric matrices.  Since $T$ is orthogonal, we see that 
we have essentially computed a square root for (\ref{oa-hamiltonian}).  
We have explicitly found the $B$ field and the $e^\sigma$ term is 
roughly the square root of $ds^2$.  This shows that at a point on 
$\oaftm$ the moduli space for $ds^2_{\cal H}$ is given by $O_Q(2n)/K$.  
The moduli space depends on $n^2$ real parameters.  The reader should 
compare this with the standard discussion in toroidal target spaces.

It is now worthwhile to consider a second decomposition which is 
reminiscent of both the Iwasawa decomposition and the Euler angle 
decomposition.  Observe that there exists an $R\in O(n)$ such that 
$\sigma = R^{-1}\Delta R$ where $\Delta$ is diagonal.  What we do is 
insert this into (\ref{oa-decomp-c}), absorb an $R^{-1}$ into $T$, 
redefine $B$ and push $R$ all the way to the right.
\begin{equation}
    T'
    \left(
    \begin{array}{cc}
        e^{+\Delta} & 0  \\
        0 & e^{-\Delta}
    \end{array}
    \right)
    \left(
    \begin{array}{cc}
        I & 0  \\
        -B' & I
    \end{array}
    \right)
    \left(
    \begin{array}{cc}
        R & 0  \\
        0 & R
    \end{array}
    \right)
    \left(
    \begin{array}{c}
        \omega \\
        \mu 
    \end{array}
    \right)
    =   T'
    \left(
    \begin{array}{c}
         e^{+\Delta} \omega' \\
        e^{-\Delta}(\mu' - B'\omega')
    \end{array}
    \right) 
    \;.
\end{equation}
In the above $T'\in K$, $B'$ is skew, $\omega'=R\omega$ and 
$\mu'=R\mu$.  Many explicit examples of target space duality directly 
give a hamiltonian which can be put into this final form.  This 
decomposition depends crucially on the details of how the subgroup $K$ 
sits inside of $O_Q(2n)$.

\begin{figure}[tbp]
    \centerline{\epsfysize 0.35\textheight \epsfbox{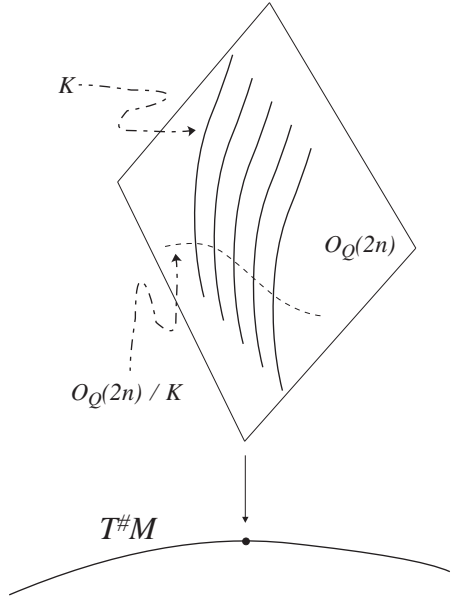}} 
    \caption[bundle]{The quadrilateral represents the fiber $O_Q(2n)$ 
    over a point in $\oaftm$.  The coset space $O_Q(2n)/K$ arises from 
    a fibration of $O_Q(2n)$ with the fibers isomorphic to $K$.  The 
    leaves of the fibration are depicted by the heavy vertical curves.  
    We schematically denote the orbit space $O_Q(2n)/K$ by the dashed 
    curve.} \protect\label{oa-fig-bundle}
\end{figure}

The symplectic structure on the infinite dimensional phase space is 
associated with $ds^2_Q$.  This gives us a frame bundle with fiber 
$O_Q(2n)$ as in Figure~\ref{oa-fig-bundle}. Fix a point in $\oaftm$.
Each orbit of $K$ in $O_Q(2n)$ determines a hamiltonian at that point.  
A choice of a $K$-orbit on the fiber over each point in $\oaftm$ is 
same as a choice of a sub-bundle with reduced structure group $K$.  In 
general this bundle will not be associated with a hamiltonian.  
Remember that earlier we were careful to state that the moduli space 
for the hamiltonian {\em at a point} in $\oaftm$ was $O_Q(2n)/K$.  If 
$(x,p)$ are local coordinates on $\oaftm$ then observe that 
$g_{\mu\nu}(x)$ and $B_{\mu\nu}(x)$ are only functions of $x$.  They 
do not depend on $p$.  Bundles of type $K(\oaftm,ds^2_Q,ds^2_{\cal 
H})$ are a subset of the set of bundles of $K$-orthonormal frames.  
This has important consequences when studying the principal 
$K$-bundles over $\oaftm$ because the bundles 
$K(\oaftm,ds^2_Q,ds^2_{\cal H})$ are very special.  If one writes the 
structural equations for a general principal $K$-bundle one finds many 
curvature and torsion terms.  The fact that $ds^2$ and $B$ ``come from 
$M$'' imposes many constraints on the structural equations which are 
discussed in \cite{oa-inprep}.  The structural equations contain the 
information necessary to understand the local obstructions to the 
existence of $K$-isometries.

\section{A challenge to mathematicians}

The content of these lectures suggest a very interesting mathematical 
problem.  Can one find examples of geometrical data which would lead 
to nontrivial {\em local} $K$-isometries between $\oaftm$ and 
$\oaftmt$?  The reason for the requirement of locality is that in 
these lectures we did not discuss global issues except in the case of 
toroidal target spaces.  The existence of local isomorphisms between 
the principal bundles $K(\oaftm,ds^2_Q,ds^2_{\cal H})$ and 
$K(\oaftmt,\oamett_Q,\oamett_{\cal\widetilde{H}})$ which preserve the 
canonical 1-forms is intimately related to Cartan's problem
\index{equivalence problem} of equivalence~\cite{oa-gardner}.  There one 
finds that the existence of such local isomorphisms is codified in the 
curvature and torsion of the bundles.  Explicit formulas will appear 
in~\cite{oa-inprep}.  The only explicit examples known of this phenomenon 
are the toroidal target spaces.  It is not clear whether the 
non-abelian duality examples fall in this ``restricted'' class.  
Global existence theorems are much more complicated~\cite{oa-gromov}.

There are well established procedures for constructing the dual theory 
when the sigma model admits a continuous group 
action~\cite{oa-G-P-R,oa-dlO-Q,oa-A-AG-L1,oa-G-R2,oa-R-V}.  Is there a 
systematic procedure which can be invoked to look for the existence of 
dual target spaces when there are no symmetries?

\section*{Acknowledgments}

I would like to thank L.~Baulieu, V.~Dotsenko, V.~Kazakov, M.~Picco 
and P.~Windey for the terrific job they did in organizing a very 
stimulating and dynamic summer institute at Carg\`{e}se.  I would like 
to thank J.~Sanchez-Guillen and the Theoretical Physics Group of the 
University of Santiago for their hospitality during my visit.  I am 
indebted to the Iberdrola Foundation for the visiting professorship 
which made my visit to Santiago possible.  This research was supported 
in part by The National Science Foundation, Grant PHY--9507829.

Many colleagues have contributed to my understanding of target space 
duality.  I am most indebted to Robert Bryant for explaining to me the 
many mysteries of exterior differential systems and Cartan's method of 
equivalence.  He has patiently and generously answered my questions.  
Is Singer always clears my mathematical confusions with clear and 
succinct explanations.  Chien-Hao Liu has been a fantastic sounding 
board, always asking important probing questions which have helped 
clarify my ideas.  I would like to thank Thom Curtright for sharing 
some of his ideas with me.  Joaquin Sanchez-Guillen patiently listened 
and provided feedback during our many discussions at Santiago.  I 
would like to thank Paul Windey for many discussions at conception of 
this program for studying target space duality.  I would like to thank 
Chien-Hao Liu and Paul Watts for their critical comments on this 
manuscript.


\begin{thebibliography}{MM}

\bibitem{oa-G-R-V}
A.  Giveon, E.  Rabinovici and G.  Veneziano, ``Duality in string
background space'', Nucl.  Phys.  B322 (1989) 167.

\bibitem{oa-C-Z} T. Curtright and C. Zachos,
  ``Currents, charges, and canonical structure of pseudodual chiral
  models'', Phys. Rev. D49 (1994)  5408 - 5421, {\tt hep-th/9401006}.

\bibitem{oa-A-AG-L2}
E.  Alvarez, L.  Alvarez-Gaum\'e and Y.  Lozano, ``A canonical approach
to duality transformations'',  Phys. Lett. B336 (1994) 183-189, {\tt
hep-th/9406206}.

\bibitem{oa-Lo} Y. Lozano,
  ``Non-abelian duality and canonical transformation'', 
  Phys.Lett. B355 (1995) 165-170,
{\tt   hep-th/9503045}.


\bibitem{oa-G-P-R} A. Giveon, M. Porrati and E. Rabinovici,
  {\it Target space duality in string theory},
  Phys. Reports 244 (1994)   77 - 202, {\tt hep-th/9401139}.

\bibitem{oa-dlO-Q}
X.  de la Ossa and F.  Quevedo, ``Duality symmetries from nonabelian
isometries in string theory'', Nucl.  Phys.  B403 (1993) 377, {\tt
hep-th/9210021}.

\bibitem{oa-F-J}
B.E.  Fridling and A.  Jevicki, ``Dual representations and ultraviolet
divergences in nonlinear sigma models'', Phys.  Lett.  134B (1984) 70.


\bibitem{oa-A-AG-L1}
E. Alvarez, L. Alvarez-Gaum\'e and Y. Lozano,
``On Non-abelian Duality'', Nucl. Phys. B424 (1994) 155-183,
{\tt hep-th/9403155}.

\bibitem{oa-G-R2}
A. Giveon and M. Ro\u{c}ek, ``On Nonabelian Duality'',
Nucl. Phys. B421 (1994)  173, {\tt hep-th/9308154}.


\bibitem{oa-G-R3} A. Giveon and M. Ro\u{c}ek,
  ``Introduction to duality'',
  {\tt hep-th/9406178}.

\bibitem{oa-R-V}
M.  Ro\u{c}ek and E.  Verlinde, ``Duality, quotients, and currents'', Nucl.
Phys.  B373 (1992) 630, {\tt hep-th/9110053}.

\bibitem{oa-yau} Yau, S.T., {\it Essays on Mirror Manifolds}, edited by 
S.T.~Yau, International Press, 1992.

\bibitem{oa-gardner} R.B.~Gardner, ``The method of equivalence and its 
applications'', CBMS-NSF Regional Conference Series in Applied 
Mathematics, Vol.~58, SIAM, 1989.

\bibitem{oa-inprep} O. Alvarez, ``The classical geometry of target space 
duality'', Part I and Part II, in preparation.

\bibitem{oa-B-C-G3} R.~Bryant, S.S.~Chern, R.B.~Gardner, 
H.L.~Goldschmidt, P.A.~Griffiths, {\it Exterior Differential Systems\/}, 
Mathematical Sciences Research Institute publication~18, 
Springer-Verlag, 1991.

\bibitem{oa-A-M} R. Abraham and J.E. Marsden,
  {\it Foundations of mechanics}, 2nd ed., Benjamin/Cummings Publ.,
  1978.

\bibitem{oa-Ar} V.I. Arnold,
  {\it Mathematical methods of classical mechanics}, 2nd ed., GTM~60,
  Springer-Verlag, 1989.

\bibitem{oa-K-S1} C.~Klim\v{c}\'{\i}k and P.~\v{S}evera,
``Strings in spacetime cotangent bundle and T-duality'',
Mod. Phys. Lett. A10 (1995)  323-330,
{\tt hep-th/9411003}.

\bibitem{oa-K-S2} C.~Klim\v{c}\'{\i}k and P.~\v{S}evera,
``Dual and non-abelian duality and the Drinfeld double'',
Phys. Lett. B351 (1995) 455-462, {\tt hep-th/9502122}.

\bibitem{oa-gromov} M.~Gromov, {\em Partial Differential Relations\/}, 
Springer-Verlag, 1986.

\bibitem{oa-A-AG-B-L}
E.~Alvarez, L.~Alvarez-Gaum\'{e}, J.~Barb\'{o}n and Y.~Lozano,
``Some global aspects of duality in string theory'', 
Nucl. Phys. B415 (1994)  71-100, {\tt hep-th/9309039}.
\end{thebibliography}
\end{document}